\documentclass[fleqn,twoside]{article}

\usepackage{espcrc2}
\usepackage{graphicx}
\usepackage[usenames]{color}
\usepackage[figuresright]{rotating}

\newcommand{\kval}{\kappa_{\rm val}}
\newcommand{\ksea}{\kappa_{\rm sea}}

\newcommand{\AmS}{{\protect\the\textfont2
  A\kern-.1667em\lower.5ex\hbox{M}\kern-.125emS}}

\definecolor{yellow}{rgb}{1.00, 1.00, 0.00}
\definecolor{red}{rgb}{1.00, 0.00, 0.00}
\definecolor{darkgreen}{cmyk}{1,0,1,0.5}
\definecolor{darkblue}{cmyk}{1,1,0,0.5}
\definecolor{Black}{rgb}{0.00, 0.00, 0.00}

\title{An analysis of the vector meson spectrum from lattice QCD}
\author{W.~Armour
{\thanks{The author wishes to thank PPARC for funding. \mbox{C. Allton} for his guidance, help and contributions, also the Adelaide group for there useful comments.}
}
\\Department of Physics, University of Wales Swansea, Swansea, Wales, SA2~8PP
}

\begin{document}


\begin{abstract}
We re-analyse meson sector data from the CP--PACS collaboration's dynamical simulations \cite{cppacs}. Our analysis uses several different approaches, and compares the standard n\"aive linear fit with the Adelaide Anzatz.
We find that setting the scale using the J parameter gives remarkable agreement among data sets. Our predictions for the $\rho$ and $\phi$ masses have very small statistical errors, $\sim 3$ MeV, but the discrepancy between the different fitting approaches is $\sim 40$ MeV.
\end{abstract}



\maketitle



\section{Introduction}
\begin{figure}[htb]
\includegraphics[width=17.5pc]{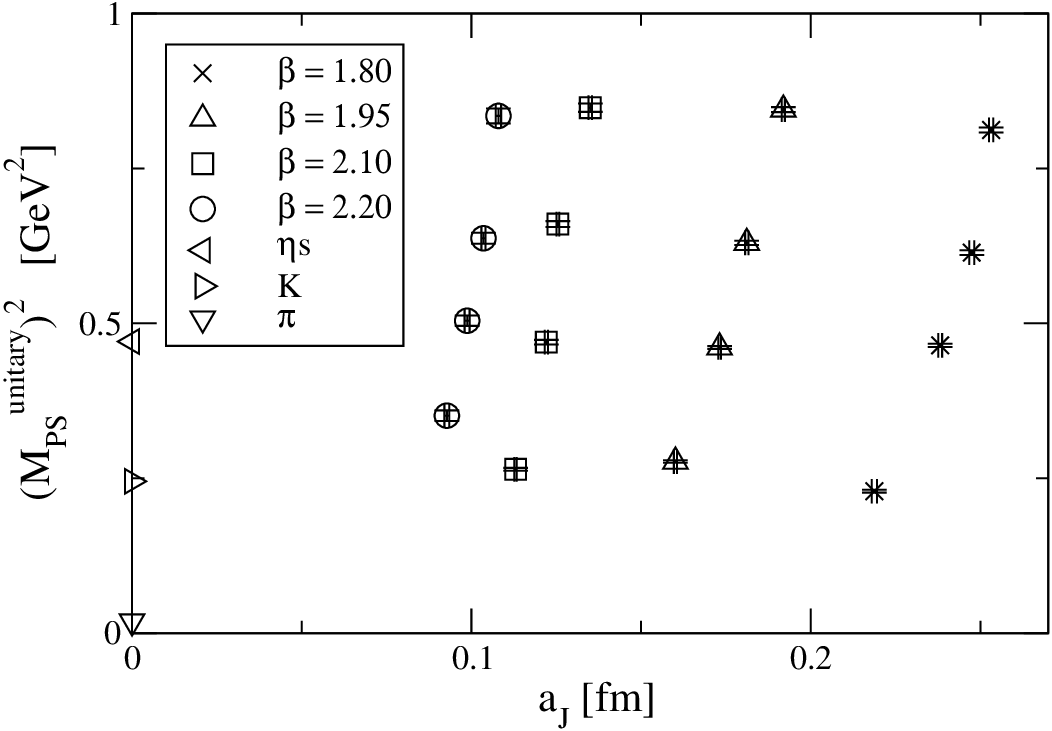}
\vspace{-8mm}
\caption{A Graphical overview of the CP--PACS data. Note that here the scale is set using the inverse lattice spacing determined from the J parameter \cite{leonardo}.}
\label{fg:avsmps}
\vspace{-5mm}
\end{figure}

We study the chiral extrapolation of the vector meson data from CP--PACS \cite{cppacs}. We do not have access to CP-PACS original data, so we produced Gaussian distributions with central values and FWHM's equal to the quoted central values and errors respectively. Our data is uncorrelated throughout and we only fit to degenerate data, i.e. $\kval^1 = \kval^2$.
The CP--PACS data used is from mean--field improved Wilson fermions with improved glue at four different $\beta$ values. For each $\beta$ value there are four different $\ksea$ values, giving sixteen independent ensembles.The physical volume was held fixed at $La \approx 2.5$ fm for the $\beta=1.80, 1.95$ and $2.10$, but the $\beta=2.20$ ensemble had a slightly smaller physical volume. A graphical overview of the CP--PACS data is given in Figure \ref{fg:avsmps}. We also fit to CP--PACS quenched data for comparison. 
%
%
%



\section{Fitting analysis}
\subsection{Summary of analysis techniques}
Our chiral extrapolation approach is based upon converting all masses into physical units prior to any extrapolation being performed. An alternative approach would be to extrapolate dimensionless masses (in lattice units) \cite{cppacs}.
Our method has the following two advantages:
\begin{itemize}
\item Different ensemble's data can be combined together in a global fit.
\item Dimensionful mass predictions from lattice simulations are effectively mass ratios, and hence one would expect some of the systematic and statistical errors to cancel.
\end{itemize}
\subsection{Fitting functions}
In our chiral extrapolations we use the n\"aive linear fit, Eq.(\ref{eq:linear}), as well as the Adelaide Anzatz, \cite{adel_rho} Eq.(\ref{eq:adel}).
\begin{eqnarray} \label{eq:linear}  \nonumber
M_V(\beta,\ksea;\kval,\kval) =
\end{eqnarray}
\vspace{-8mm}
\begin{equation}
C_0 + C_2 M^2_{PS}(\beta,\ksea;\kval,\kval).
\end{equation}
\begin{eqnarray} \label{eq:adel} \nonumber 
M_V(\beta,\ksea;\kval,\kval) = 
\end{eqnarray}
\vspace{-8.5mm}
\begin{eqnarray} \nonumber
C_0 + C_2 M^2_{PS}(\beta,\ksea;\kval,\kval)~+
\end{eqnarray}
\vspace{-7mm}
\begin{eqnarray} 
\frac{1}{2}\frac{\left[ \Sigma^\rho_{\pi\pi}+ \Sigma^\rho_{\pi\omega}\right]}
{(C_0 + C_2 M^2_{PS}(\beta,\ksea;\kval,\kval)}
\end{eqnarray}
where the Self Energies are defined as $\Sigma^\rho_{\pi\pi}=\Sigma^\rho_{\pi\pi}(M^2_{PS}(\beta,\ksea;\ksea,\kval))$ and $\Sigma^\rho_{\pi\omega}=\Sigma^\rho_{\pi\omega}(M^2_{PS}(\beta,\ksea;\ksea,\kval))$.
$M_{V(PS)}$ is the vector(pseudo-scalar) meson mass.
$\beta$ and $\ksea$ refer to the sea structure (i.e. the gauge coupling and sea quark hopping parameter) and $\kval$ refers to the valence quark hopping parameters.
\\
\\
The Adelaide method relies on a parameter $\Lambda_{\pi\omega}$ \cite{adel_rho}. We use the value of $\Lambda_{\pi\omega} = 630$ MeV taken from \cite{adel_rho}. The value of $\Lambda_{\pi\omega}$ is highly constrained by the {\em lightest} data point in the $M_V$ versus $M_{PS}^2$ plot, and since the data used in \cite{adel_rho} includes a much lighter point than in this study, we use its value of $\Lambda_{\pi\omega}$.
\subsection{Individual ensemble fits}

\begin{figure}[htb]
\vspace{-4mm}
\includegraphics[width=17.5pc]{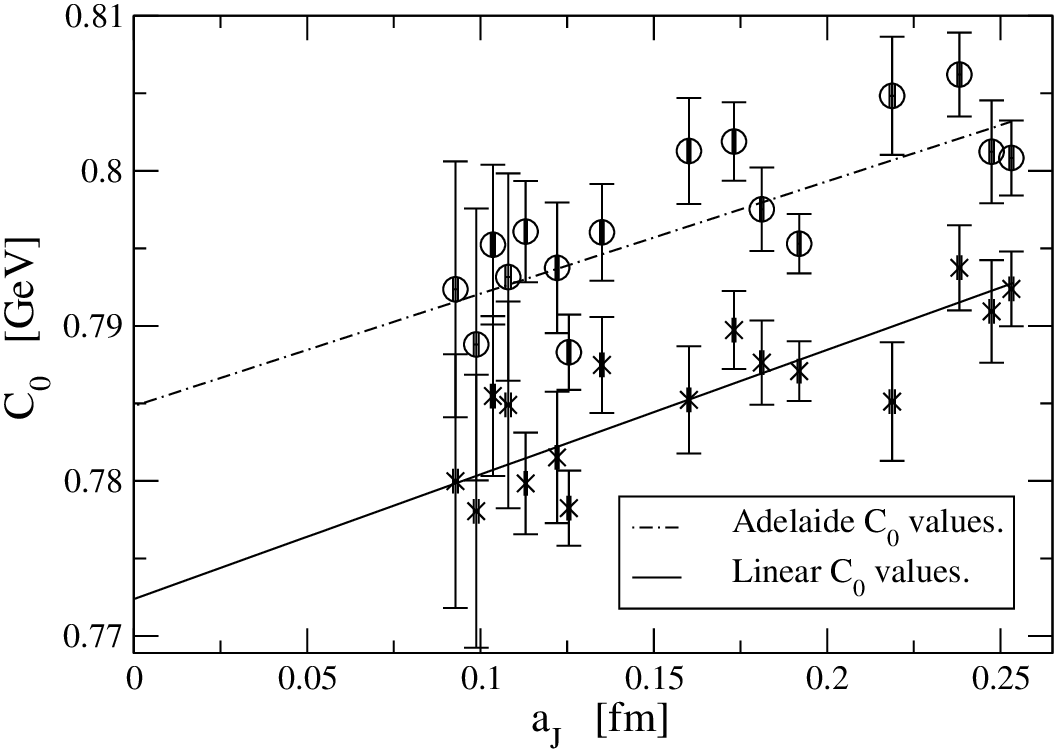}
\vspace{-8mm}
\caption{An example of a continuum extrapolation performed to fitting parameters obtained from the sixteen different data sets.}
\label{fg:C0_a}
\vspace{-5mm}
\end{figure}

Our first method for obtaining physical masses involves determining fitting parameters, $C_0$ and $C_2$, for each of the 16 data sets and then performing a continuum extrapolation to those. Figure \ref{fg:C0_a} displays the $C_0$ values and motivates a continuum extrapolation of the form in Eq.(\ref{eq:C02_cont}).
%
%
%
%
\begin{equation} \label{eq:C02_cont} 
C_{0,2}(a) = C_{0,2}^{cont} + X_{0,2} \;a_J.
\end{equation}
Table 1 lists the results for these fits.
%
%
\begin{table*}[htb]
\label{tb:C02_cont}
\begin{tabular}{@{}lcccccc@{}}
\hline
  & $C_{0}^{cont.}$ & $X_{0}$ & $\chi^{2}_{0}/d.o.f.$ & $C_{2}^{cont.}$ & $X_{2}$         &  $\chi^{2}_{2}/d.o.f.$ \\
  & [GeV]           & [GeV/fm]&                       & [GeV$^{-1}$]    & [GeV$^{-1}$/fm] & \\
\hline
Linear-fit    & $0.772\pm{ 3}$ & $8\pm{ 2}\times 10^{ -2}$ & $1.10/14$ & $0.473\pm{ 6}$ & $-0.27\pm{ 3}$ & $1.86/14$ \\ 
Adelaide-fit  & $0.785\pm{ 3}$ & $7\pm{ 2}\times 10^{ -2}$ & $2.01/14$ & $0.462\pm{ 6}$ & $-0.26\pm{ 3}$ & $1.39/14$ \\
\hline
\end{tabular}
\vspace{1mm}
\\Table 1: The coefficients obtained from the continuum extrapolation of the parameters obtained from the 16 \emph{individual} fits using Eq.(3).
\end{table*}
%
%
\subsection{Global fits}
\begin{figure}[htb]
\vspace{-5mm}
\includegraphics[width=17.5pc]{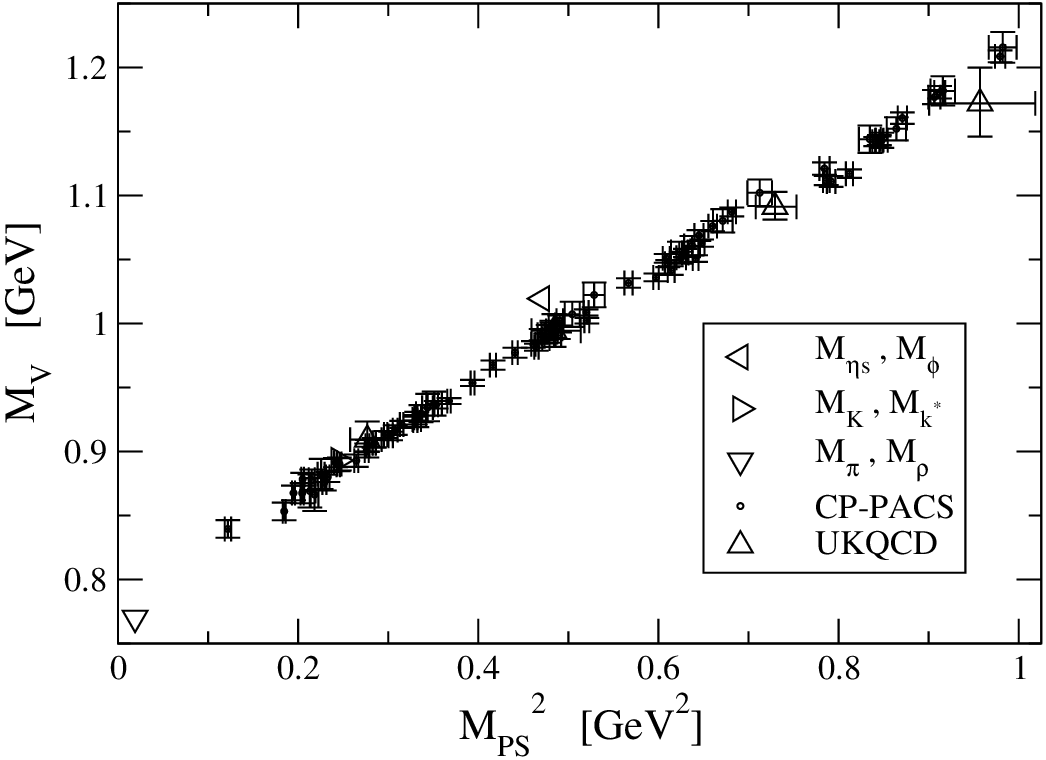}
\vspace{-8mm}
\caption{A plot of $M_V$ versus $M_{PS}^2$ for the degenerate \mbox{CP-PACS} dataset showing that the data lies on a near universal line. Also plotted are the three experimental data points, and the {\em unitary} data from UKQCD \cite{ukqcd_csw202}. Note that the scale for the lattice data was set using the method of \cite{leonardo}.}
\label{fg:global}
\vspace{-5mm}
\end{figure}

\begin{figure}[htb]
\vspace{-3mm}
\includegraphics[width=17.5pc]{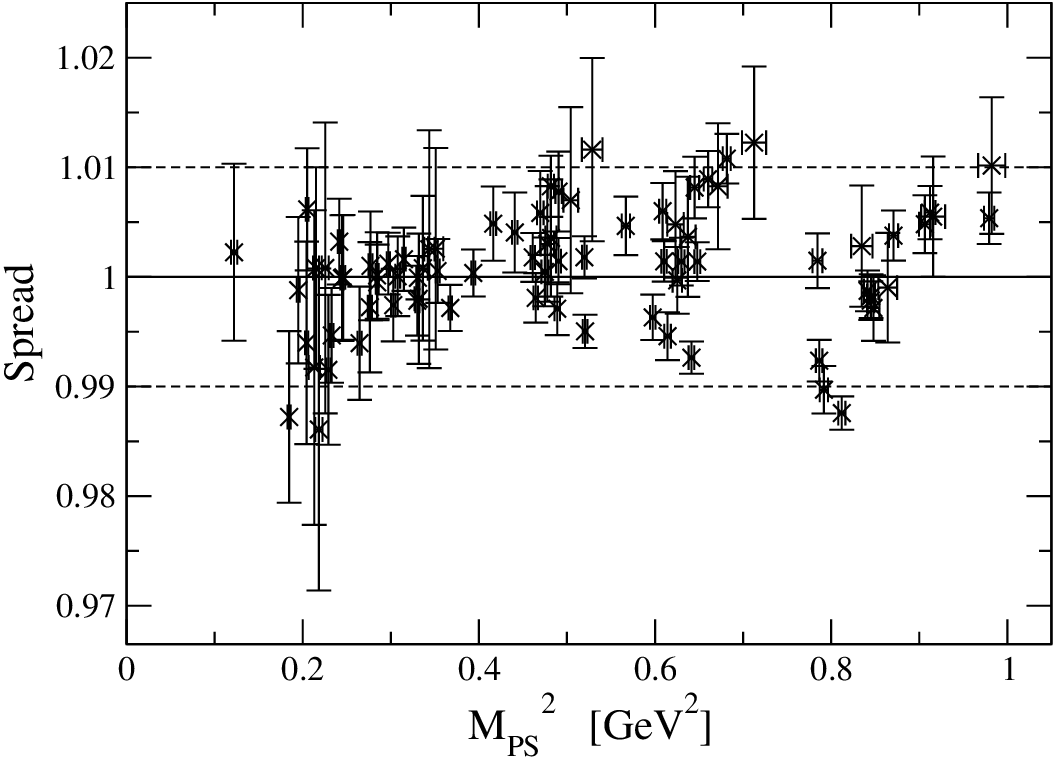}
\vspace{-8mm}
\caption{A plot showing the relative spread in $M_V$ versus $M^2_{PS}$ for the degenerate CP--PACS data set.}
\label{fg:spread}
\vspace{-5mm}
\end{figure}

Figure \ref{fg:global} plots the vector meson mass against the pseudo--scalar mass squared (in physical units) for all of the data from all of the ensembles.
As can be seen the data lies on a near universal line. This motivates an analysis which combines all of the degenerate data into one global fit.
%
%
%
%
\\
To underline the near universal nature of the data we plot the relative spread of the data using Eq.\ref{eq:spread}.
\begin{equation} 
\mbox{spread} = \frac{M_V^{data}}{C_0 + C_2 M_{PS}^2}
\label{eq:spread}
\end{equation}
Where $C_0 + C_2 M_{PS}^2$ is a simple fit through the data in Figure 3.
In Figure \ref{fg:spread} we see that the spread is at the 1\% level or less. Note that this simple linear fit is not used in any further analysis. Also note that the scale is set from J \cite{leonardo} and hence the data must pass through the (M$_K*$ , M$_K$) point.
%
%
%
%
When grouping the different ensembles together we must allow for some (small) variation with the lattice spacing. We do this by modifying the linear and Adelaide fitting functions. By substituting Eq.(\ref{eq:sub}) into Eqs.(\ref{eq:linear} \& \ref{eq:adel}) respectively. Table 2 lists the results of these fits.
\begin{equation} \label{eq:sub}
C_{0,2} \rightarrow C'_{0,2} + X'_{0,2} a.
\end{equation}
%
%
\begin{table*}[htb]
\label{tb:global}
\begin{tabular}{@{}lccccc@{}}
\hline
 & $C'_{0}$ & $X'_{0}$ & $C'_{2}$     & $X'_{2}$        & $\chi^2/d.o.f.$ \\
 & [GeV]    & [GeV/fm] & [GeV$^{-1}$] & [GeV$^{-1}$/fm] & \\
\hline
Linear--fit    & $0.772\pm{ 3}$ & $8\pm{ 2}\times 10^{ -2}$   & $0.474\pm{ 6}$ & $-0.28\pm{ 3}$ & $43/76$ \\
Adelaide--fit  & $0.784\pm{ 3}$ & $7.8\pm{16}\times 10^{ -2}$ & $0.466\pm{ 6}$ & $-0.28\pm{ 3}$ & $58/76$ \\
\hline
\end{tabular}
\vspace{1mm}
\\Table 2: The coefficients obtained from a \emph{global} fit of all the $M_V$ data against $M^2_{PS}$ using the linear-O(a), Eq.(1), and Adelaide-O(a), Eq.(2), fits, both incorporating Eq.(5). 
\end{table*}
%
%


\section{Conclusions}

To conclude, Table \ref{tb:mass_estimates} lists our mass estimates for the $\rho$ and $\phi$ mesons. We see that our interpretation of the Adelaide Anzatz underestimates M$_\rho$ presumably due to a poorly tuned $\Lambda_{\pi\omega}$ value. We also note the following:

\begin{itemize}

\item Setting the scale using J gives remarkable agreement among data sets.

\item The (statistical) errors in the mass estimates are tiny.

\item The discrepancies between the various fitting procedures is much larger than the statistical errors listed.

\item Note that for the global linear fit that incorporates the $Xa$ correction, we obtain an $M_\rho$ only 10 MeV above the experimental value.

\item The estimates of $M_\rho$ and $M_\phi$ from this approach are closer to the corresponding experimental values than the quenched estimates.

\end{itemize}


\begin{table}[hbt]
\vspace{-2mm}
\begin{tabular}{@{}lcc@{}}
\hline
  & M$_{\rho}$ & M$_{\phi}$ \\
  & [GeV]      & [GeV]      \\
\hline
Experiment & 0.770 & 1.0194  \\
Quenched + X$_{0,2}$ & $0.798\pm{ 4}$ & $0.988\pm{ 5}$ \\
Global--Linear + X$_{0,2}$ & $0.781\pm{ 3}$ & $0.995\pm{2}$ \\
Global--Adelaide + X$_{0,2}$ & $0.740\pm{ 3}$ & - \\
\hline
\end{tabular}
\vspace{1mm}
\\Table 3: Mass predictions.
\label{tb:mass_estimates}
\end{table}





\end{document}